# Speech Enhancement in Adverse Environments Based on Non-stationary Noise-driven Spectral Subtraction and SNR-dependent Phase Compensation

Md Tauhidul Islam[a], Asaduzzaman[b], Celia Shahnaz[b,*], Wei-Ping Zhu[c], M. Omair Ahmad[c]

[a]*Department of Electrical and Computer Engineering, Texas A&M University, College Station, Texas, USA-77840*
[b]*Department of Electrical and Electronic Engineering, Bangladesh University of Engineering and Technology, Dhaka-1000, Bangladesh*
[c]*Department of Electrical and Computer Engineering, Concordia University, Montreal, Quebec H3G 1M8, Canada*

**Abstract**

A two-step enhancement method based on spectral subtraction and phase spectrum compensation is presented in this paper for noisy speeches in adverse environments involving non-stationary noise and medium to low levels of SNR. The magnitude of the noisy speech spectrum is modified in the first step of the proposed method by a spectral subtraction approach, where a new noise estimation method based on the low frequency information of the noisy speech is introduced. We argue that this method of noise estimation is capable of estimating the non-stationary noise accurately. The phase spectrum of the noisy speech is modified in the second step consisting of phase spectrum compensation, where an SNR-dependent approach is incorporated to determine the amount of compensation to be imposed on the phase spectrum. A modified complex spectrum is obtained by aggregating the magnitude from the spectral subtraction step and modified phase spectrum from the phase compensation step, which is found to be a better representation of enhanced speech spectrum. Speech files available in the NOIZEUS database are used to carry extensive simulations for evaluation of the proposed method.

*Keywords:* Speech enhancement, spectral subtraction, magnitude compensation, phase compensation, noise estimation

## 1. Introduction

Corruption of speech signals by the additive or multiplicative noise deteriorates the performance of speech processing applications such as speech communication, speech recognition etc. Removing the disturbing noise while preserving the speech is desirable for proper operation of these systems. Many speech enhancement methods have been proposed to achieve this goal. Spectral subtraction [1, 2, 3, 4, 5], Wiener filter [6, 7], minimum mean square error (MMSE) estimator [8, 9], subspace based methods [6, 7], thresholding methods based on wavelet transform [10, 11, 12, 13] and Kalman filtering [14] are the prominent ones. Subspace and wavelet based approaches are computationally slow. Frequency-domain methods are computationally fast, but most of them need an estimation of noise to perform speech enhancement.

Low computation with good performance in stationary noise makes spectral subtraction a very attractive and widely used method. In spectral subtraction, the noise spectrum is estimated and subtracted from the noisy speech spectrum. If there is no variation of the noise with time which means that the noise is stationary, the method works well. However, in presence of non-stationary noise, the performance of this method degrades because of its inability to estimate the noise properly. Another problem of this method is the presence of musical noise, a noise of increasing variance. The first problem is the main concern of [15], where noise is estimated based on the high-order Yule-Walker equations without finding the non-speech frames. This method can track the non-stationary noise but computationally exhaustive. Another method named minimum statistics based spectral subtraction [16], can estimate the non-stationary noise with less computation but this method depends on the noise estimation in the past frames which sometimes invokes wrong estimates or leads to speech distortion. In [17], noise spectrum is estimated based on information of the high frequency spectrum of the current frame. This method requires very high sampling rate which creates significant problems in the context of speech processing application.

*Corresponding author
 Email address: celia.shahnaz@gmail.com (Celia Shahnaz)

In the above mentioned methods, although the spectrum of the noisy speech is a complex number, only the magnitude is modified based on the estimate of the noise spectrum and phase remains unchanged. This was being done for a long time based on an assumption that human auditory system is phase-deaf, i.e., cannot differentiate change of phase, until the authors in [18] showed that the phase spectrum can also be very useful in speech enhancement. The authors used the phase spectrum in a spectral subtraction based approach to obtain an enhanced speech. Later, the authors in [19, 20] also used this idea for speech enhancement. But these methods did not consider the magnitude spectrum at all. In this paper, we consider both magnitude and phase spectra and compensate both of them based on the noise characteristics. We develop a noise estimation approach that can track the time variation of non-stationary noise for magnitude spectrum compensation. The phase spectrum is compensated in an SNR-dependent phase compensation step. We aggregate the modified magnitude and phase from these two steps and we find this modified complex spectrum effective in producing enhanced speech of improved quality with minimal speech distortion as compared to some of the state-of-the-art speech enhancement methods.

The paper is organized as follows. Section 2 presents the proposed method. Section 3 describes results. Concluding remarks are presented in section 4.

## 2. Problem Formulation and Proposed Method

In any analysis, modification and synthesis (AMS) framework, at first, noisy speech frames are transformed by a transformation method. Then modifications are carried out in the transformed domain and finally, the inverse transform of the transformation method followed by the overlap-add method is performed to reconstruct the enhanced speech. The proposed method is based on the AMS framework, where speech is analyzed, modified and synthesized frame wise.

In the presence of additive noise d[n], a clean speech signal x[n] gets contaminated and produces noisy speech y[n]. The noisy speech can be segmented into overlapping frames by using a sliding window. $\tau^{th}$ windowed noisy speech frame can be expressed in the time domain as

$$y^\tau[n] = x^\tau[n] + d^\tau[n], \tau = 1, \ldots, T, \qquad (1)$$

where $T$ is the total number of speech frames. If $Y^\tau[k]$, $X^\tau[k]$ and $D^\tau[k]$ are the short-time Fourier transform (STFT) representations of $y^\tau[n]$, $x^\tau[n]$ and $d^\tau[n]$, respectively, we can write

$$Y^\tau[k] = X^\tau[k] + D^\tau[k], \qquad (2)$$

where $k = 0, 1, 2, \ldots N - 1$, $N$ is the total number of samples in a frame. The $N$-point STFT, $Y^\tau[k]$ of $y^\tau[n]$ can be computed as

$$Y^\tau[k] = \sum_{n=0}^{N-1} y^\tau[n] e^{-\frac{j2\pi nk}{N}}. \qquad (3)$$

The Fourier transform of the noisy speech frame, $Y^\tau[k]$ is modified in the proposed method to obtain an estimate of the clean speech spectrum $X^\tau[k]$.

An overview of the proposed speech enhancement method is shown by a block diagram in Fig. 1. It is seen from Fig. 1 that Fourier transform is first applied to each input speech frame. The magnitude of the Fourier spectrum is modified in a spectral subtraction method based on non-stationary noise estimation, which we call step-1. The modified magnitude from step-1 is then combined with unchanged phase to obtain the modified complex spectrum. Using inverse fast Fourier transform (IFFT) and overlap and add, an intermediate speech signal is obtained. The spectrum of the intermediate speech is sent to step-2, which consists of phase spectrum compensation (PSC) [18]. PSC modifies the phase spectrum based on the SNR of the intermediate speech. Using the modified phase spectrum with the modified magnitude spectrum from the first step, we obtain an enhanced complex spectrum. Finally, using IFFT and overlap and add, an enhanced speech is constructed. The full AMS process is done for both steps to get full flexibilities of using different window sizes and parameters.

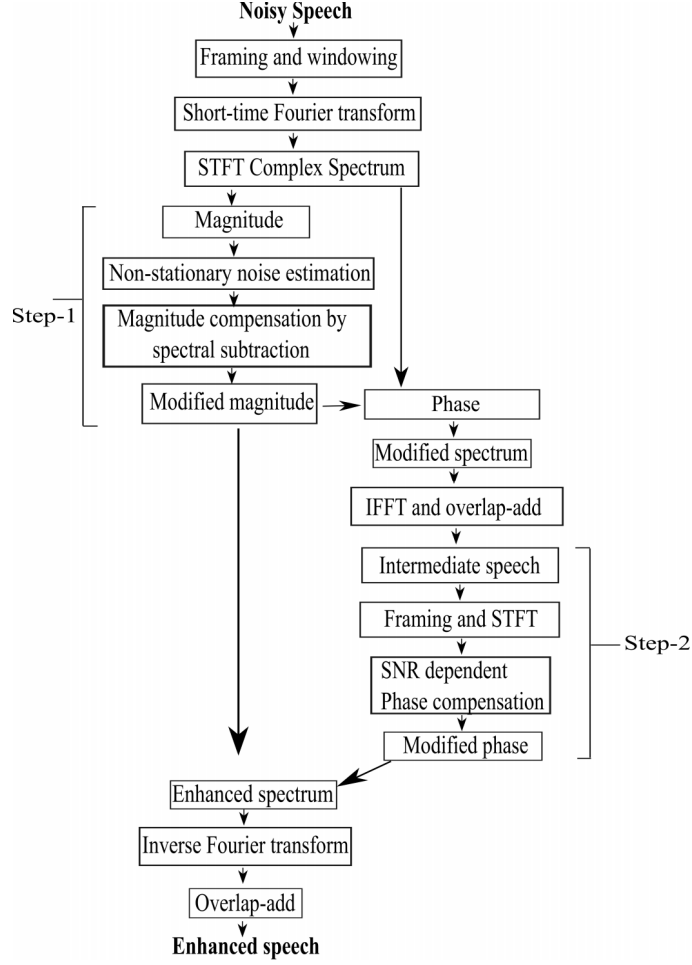

Figure 1: Block diagram of the proposed method.

*2.1. Magnitude Spectrum Compensation by Spectral Subtraction Based on Non-stationary Noise Estimation*

In this section, the magnitude spectrum of the noisy speech is modified based on the estimation of non-stationary noise. Unlike conventional methods, the estimate of noise spectrum is updated in every silence period and the low frequency region of magnitude spectrum is taken into consideration in order to compensate for the noise estimation errors that may be induced when the additive noise is non-stationary, i.e., changes its amplitude drastically with time.

We proposed to obtain an estimate of $|X^\tau[k]|$, $|Z^\tau[k]|$ as

$$|Z^\tau[k]| = \begin{cases} H^\tau[k], & \text{if } H^\tau[k] > 0, \\ \beta_s |Y^\tau[k]|, & \text{otherwise,} \end{cases} \qquad (4)$$

where

$$H^\tau[k] = |Y^\tau[k]| - \alpha^\tau |\widehat{D^{\tau_S}}[k]|. \qquad (5)$$

In (4), $\beta_s$ refers to the spectral flow parameter introduced to prevent any negative value in $|Z^\tau[k]|$. In (5), $\alpha^\tau$ symbolizes the tracking factor which tracks the change of amplitude of the non-stationary noise spectrum with time and $|\widehat{D^{\tau_S}}[k]|$ denotes the estimated noise spectrum in previous silence frame. In the proposed spectral subtraction based noise reduction scheme, the noise spectrum estimated from the beginning silence frames is updated during each silence period as follows

$$|\widehat{D^{\tau_S}}[k]| = \begin{cases} \frac{|Y^1[k]| + \cdots + |Y^{N_s}[k]|}{N_s}, & \text{for } \tau_S = 1, \\ v_n |D^{\tau_P}[k]| + (1 - v_n)|Y^{\tau_S}[k]|, & \text{otherwise,} \end{cases} \qquad (6)$$

where $N_s$ is the number of initial silence frame, $\tau_P$ refers to the index of the previous silence frame with respect to $\tau_S$ and $v_n$ is the forgetting factor. Considering that this estimate of the noise power spectrum is updated only during a silence period while it may change drastically with time, it is insufficient to use a constant value of the tracking factor $\alpha^\tau$ to compensate for the errors induced in the noise spectrum to be subtracted from the noisy speech spectrum at each frame. In order to track the time variation of the noise, $\alpha^\tau$ should be adjusted at each frame after a silence period. According to the spectral characteristics of human speech, the low frequency band typically from 0 to 50 Hz contains no speech information. Thus, for noisy speech, the low frequency band, say $\Delta = [0, 50]$ Hz contains only noise. In view of this fact, in order to change the value of $\alpha$ for $\tau^{th}$ frame, we propose to use the ratio of $|Y^\tau[k]|$ and $|\widehat{D}^{\tau_S}[k]|$ in low frequency band delta as

$$\alpha^\tau = \mu \frac{\sum_\Delta |Y^\tau[k]|}{\sum_\Delta |\widehat{D}^{\tau_S}[k]|}, \tag{7}$$

where $\Delta = [0, 50]$ Hz, $\mu$ is a constant determined empirically. In the low frequency band $\Delta$ of the $\tau^{th}$ frame, the variation of the noisy speech spectrum is equivalent to the noise spectrum of that frame. Thus, use of $\alpha_\tau$ defined in (7) clearly serves as a relative weighing factor with respect to the estimated noise spectrum $|\widehat{D}^{\tau_S}[k]|$, leading to a reasonable tracking for the time variation of the noise if non-stationary. Please note that a voice activity detector is used in the proposed scheme from [1] for detecting the speech and silence frames.

Aggregating the modified magnitude spectrum with the unchanged phase of noisy speech, we obtain a modified complex spectrum as

$$Z^\tau[k] = |Z^\tau[k]|e^{\angle Y^\tau[k]}. \tag{8}$$

After using IFFT on $Z^\tau[k]$ and overlap and add of real part of the resulting signal, we obtain time-domain intermediate speech $z[n]$.

## 2.2. SNR Dependent Phase Spectrum Compensation

If we apply STFT on $z[n]$, we obtain $Z^t[k]$, where $t$ is the frame number for step-2. In step-2, the modified complex spectrum $Z^t[k]$ is modified in such a way that the low energy component cancel out more than the high energy components. The modified complex spectrum thus obtained is a better representation of $X^t[k]$.

$$\widehat{X}^t[k] = |Z^t[k]|e^{j\angle(Z^t[k] + \phi^t[k])}. \tag{9}$$

$z^t[n]$, $t^{th}$ frame of the intermediate speech, is a real valued signal and therefore, its FFT is conjugate symmetric, i.e.,

$$Z^t[k] = \{Z^t[N_t - k]\}^*, \tag{10}$$

where $N_t$ is the number of samples in a frame in step-2. The conjugate can be obtained as a result of applying FFT on $z^t[n]$. The conjugate arise naturally from the symmetry of the magnitude spectrum and anti-symmetry of the phase spectrum. During IFFT operation as needed for synthesis of enhanced speech, the conjugates are summed together to produce larger real valued signal. If the conjugates are modified, the degree to which they sum together can be influenced and this can be contributed constructively or destructively to the reconstruction of the enhanced time domain speech. We propose the degree of phase spectrum compensation to be dependent on the SNR estimate of the current frame thus facilitating the handling of time and frequency varying non-stationary noise conditions. For this purpose, we formulate a phase spectrum compensation function as given by

$$\phi^t[k] = \psi \Lambda[k] |\widehat{V}^t|, \tag{11}$$

where $\widehat{V}^t$ is the root mean square value of $Z^t$, where $Z^t = (Z^t[1], \ldots Z^t[N_t])^T$ [18]. In (11), $\psi$ is a real valued constant and $\Lambda[k]$ presents a weighting function expressed as

$$\Lambda[k] = \begin{cases} 1, & \text{if } 0 < \frac{k}{N} < \frac{1}{2}, \\ -1, & \text{if } \frac{1}{2} < \frac{k}{N} < 1, \\ 0, & \text{otherwise.} \end{cases} \tag{12}$$

Here, zero weighting is assigned to the values of $k$ corresponding to the non-conjugate vectors of FFT, such as $k = 0$ and $k = \frac{N}{2}$, if $N$ even.

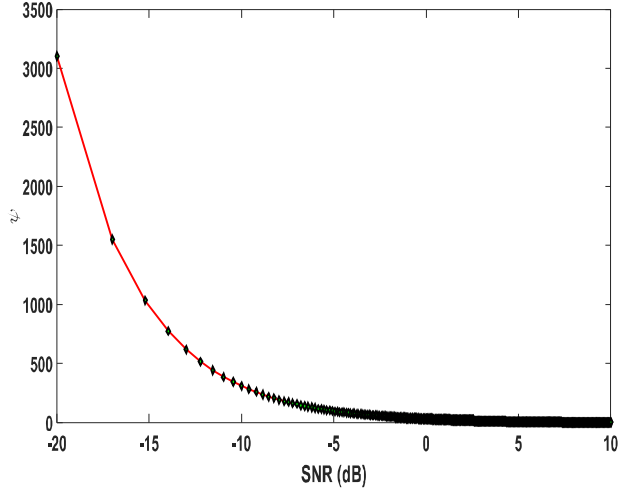

Figure 2: $\psi$ as a function of a posterior SNR.

*2.2.1. SNR Dependent $\psi$*

Unlike [18], instead of considering $\psi$ as a constant, we propose to determine it as

$$\psi = \frac{\pi^3}{\nu}, \tag{13}$$

where $\nu$ is defined as

$$\nu = \frac{|\widehat{Y^t}[k]|^2}{|\widehat{V^t}|^2}. \tag{14}$$

The right hand side of (14) is the *a posteriori* SNR of the $t^{\text{th}}$ intermediate speech frame and the plot of $\psi$ with SNR is shown in Fig. 2. It is seen from Fig. 2 that if the SNR increases, the value of the constant $\psi$ decreases and phase compensation becomes less so that there is no distortion in the signal. On the contrary, when noise increases to a higher level, $\psi$ increases. As a result, the phase compensation on the signal increases and denoising is obtained to a significant extent. Since the estimate of noise magnitude spectrum $|\widehat{V^t}|$ is constant, introduction of the weighting function $\Lambda[k]$ defined by (11) produces an anti-symmetric compensation function $\phi^t[k]$ that acts as the cause for changing the angular phase relationship in order to achieve noise cancellation during synthesis. Although detail explanation of the phase compensation method is given in [18], we revisit the explanation for clarity of our method. Explanation for two cases of single conjugate pair and their corresponding modifications, i.e., when the estimated speech vector from first step is greater and smaller than the phase compensation function are presented in Fig. 3, where both the time frequency indexes are omitted for convenience and clarity. We will denote the phase compensation function as $\phi$, the two conjugates of $Z^\tau[k]$ as $\vec{Z}$ and $\vec{Z^*}$, and of $\widehat{X^t}[k]$ as $\vec{X}$ and $\vec{X^*}$. For the representation in Fig. 3(a), the magnitude of $\vec{Z}$ and $\vec{Z^*}$ are considered larger than $\phi$. Column one of Fig. 3(a) shows the conjugate vectors $\vec{Z}$ and $\vec{Z^*}$ as well as their summation vector $\vec{Z + Z^*}$, in column two the real part of the $\vec{Z}$ and $\vec{Z^*}$ are shown to be offset by $|\phi|$ and $-|\phi|$, respectively. Altering the angles of the vectors $\vec{Z}$ and $\vec{Z^*}$ while keeping their magnitude unchanged thus produces vectors $\vec{X}$ and $\vec{X^*}$, respectively. It is seen from the column three that the vector $\vec{X + X^*}$ is produced as a result of adding the modified vectors $\vec{X}$ and $\vec{X^*}$. Column four demonstrates the real part of the addition vector $\vec{X + X^*}$, while its imaginary part is discarded with a view to avoid getting complex time domain frames after IFFT operation. Comparing column one and four of Fig. 3(a), it is clear that a limited change of original signal occurs if $|\vec{Z}|$ and $|\vec{Z^*}|$ are greater than $|\phi|$. In Fig. 3(b), similar illustration is shown by considering $|\vec{Z}|$ and $|\vec{Z^*}|$ is smaller than $|\phi|$ and found that significant change of the original signal occurs. Since $\phi$ is anti symmetric, the angle of the conjugate pair in each case of Fig. 3 are pushed in opposite directions, one towards 0 radian and other towards $\pi$ radian. The Further they are pushed apart, the more out of phase they become. This justifies that, FFT spectrum of noisy speech with larger magnitude undergoes less attenuation and that with smaller magnitude undergoes more.

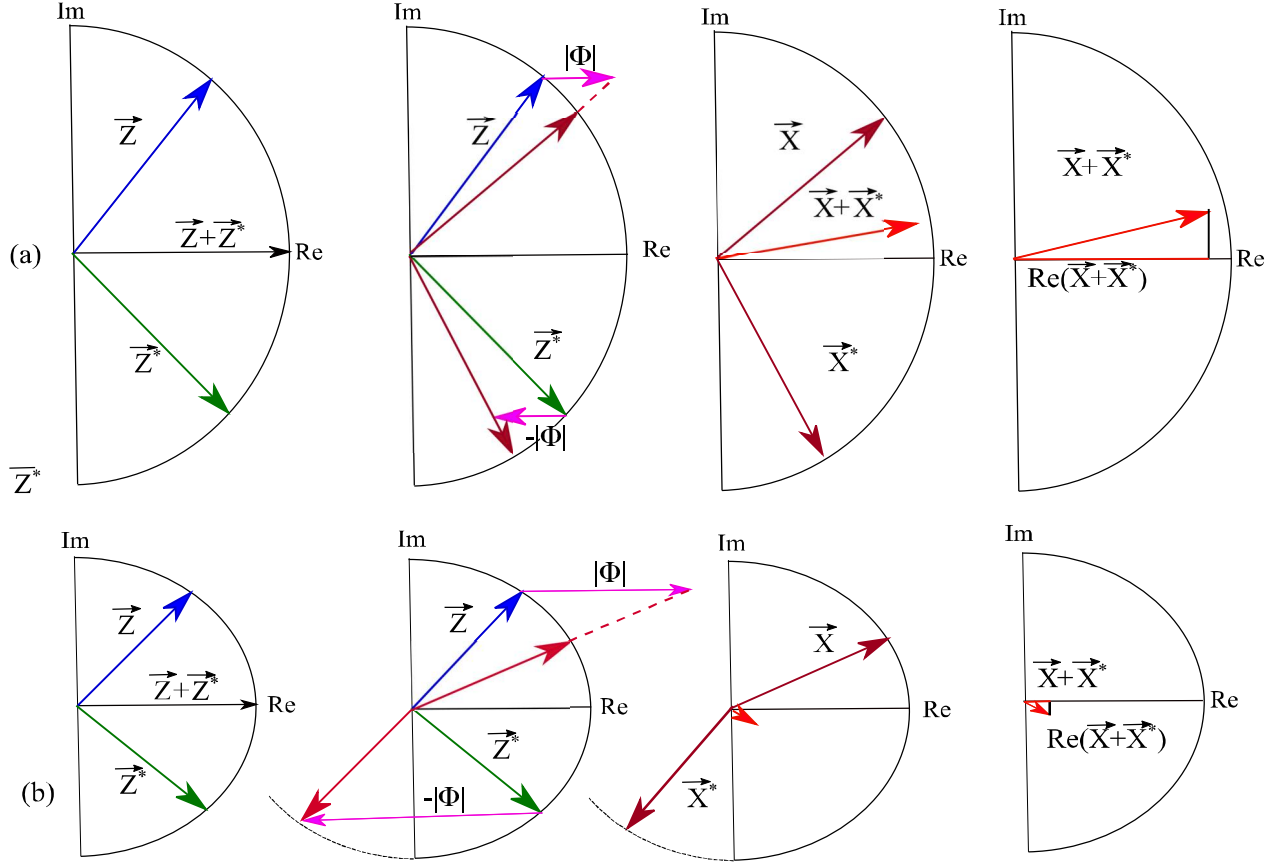

Figure 3: Phase spectrum compensation (a) when $|\vec{Z}| > |\phi|$ (b) when $|\vec{Z}| < |\phi|$.

### 2.3. Resynthesis of Enhanced Signal

The enhanced speech frame is synthesized by performing the IFFT on the resulting $\widehat{X}^t[k]$,

$$\widetilde{x}^t[n] = Re\left(IFFT\{\widehat{X}^t[k]\}\right), \tag{15}$$

where $Re(\cdot)$ denotes the real part of the number inside it and $\widetilde{x}^t[n]$ represents the enhanced speech frame. The final enhanced speech is synthesized by using the standard overlap and add method [21].

## 3. Results

In this section, a number of simulations is carried out to evaluate the performance of the proposed method.

### 3.1. Implementation

The above proposed method which we call non-stationary noise-driven spectral subtraction with SNR-dependent phase compensation (NSSP) is implemented in MATLAB R2016b graphical user interface development environment (GUIDE). The MATLAB software with its user manual is attached as supplementary material with the paper. This software also includes implementation of some recent methods, i.e., multi-band spectral subtraction (MBSS) [22], phase spectrum compensation (PSC) [18] and soft mask estimator with posteriori SNR uncertainty (SMPO) [23]. The implementations of these methods have been taken from publicly available and trusted sources. MBSS code is taken from `https://www.mathworks.com/matlabcentral/fileexchange/7674-multi-band-spectral-subtraction`, PSC implementation code is acquired from `https://www.mathworks.com/matlabcentral/fileexchange/30815-phase-spectrum-compensation` and SMPO code is taken from `http://ecs.utdallas.edu/loizou/cimplants/`. The MATLAB implementations of the calculations of segmental and overall SNR improvement are taken from `http://ecs.utdallas.edu/loizou/cimplants/` [24].

Table 1: Constants used in the spectral subtraction step

| Constant | Value |
|---|---|
| $\beta$ | 0.1 |
| $\nu_n$ | 0.167 |
| $\mu$ | 0.1 |

*3.2. Simulation Conditions*

Real speech sentences from the NOIZEUS database are employed for the experiments, where the speech data are sampled at 8 kHz. To imitate a noisy environment, noise sequence is added to the clean speech samples at different SNR levels ranging from 10 dB to −20 dB. Two different types of noises, such as babble and street are adopted from the NOIZEUS database.

In order to obtain overlapping analysis frames in the spectral subtraction step, Hamming windowing operation is performed, where the size of each of the frame is 96 samples with 50% overlap between successive frames. In the phase compensation step, Griffin and Lim's modified Hanning window is used and the size of each frame is 256 samples with 25% overlap. Values of used constants in the first step are given in Table 1.

*3.3. Comparison Metrics*

Standard Objective metrics [39], namely, segmental SNR (SNRSeg) improvement in dB, overall SNR improvement in dB, perceptual evaluation of speech quality (PESQ) are used for the evaluation of the proposed NSSP method. The proposed method is subjectively evaluated in terms of the spectrogram representations of the clean speech, noisy speech and enhanced speech. Formal listening tests are also carried out in order to find the analogy between the objective metrics and subjective sound quality. The performance of our method is compared with MBSS [22], PSC [24] and SMPO [23] in both objective and subjective senses.

*3.4. Objective Evaluation*

*3.4.1. Results for speech signals with street noise*

SNRSeg improvement, overall SNR improvement and PESQ scores for speech signals corrupted with street noise for MBSS, PSC, SMPO and NSSP are shown for a SNR range of −20 dB to 10 dB in Fig. 4, 5 and Table 2.

In Fig. 4, we see that the SNRSeg improvement for NSSP is the highest at the lowest SNR of −20 dB. The nearest SNRSeg improvement is shown by SMPO, which is almost half of the SNRSeg improvement by NSSP. With increment of the SNR, SNRSeg improvement for NSSP decreases. But at the highest SNR of 10 dB, NSSP shows an SNRSeg improvement of 4.1 dB, which is much better than MBSS, SMPO and PSC. Another interesting fact is that the SNRSeg improvement for NSSP increases monotonically with decrement of SNR. But SNRSeg improvement for other methods increase upto SNR of −5 dB, then starts to decrease. Higher SNRSeg improvement of the proposed NSSP method in all SNRs attests that NSSP can enhance the noisy speech better than other competing methods in favorable as well as adverse environments.

In Fig. 5, where we plot the overall SNR improvement for all the methods for SNR range of −20 to 10 dB, we see that NSSP provides an excellent overall SNR improvement of 14 dB at SNR level of −20 dB. Other methods provide overall SNR improvement of 11, 9 and 8.2 dB at that SNR level. NSSP continues to provide higher overall SNR improvement upto 0 dB, from where to 10 dB SNR, it provides competitive improvements in comparison to PSC and better than SMPO and MBSS.

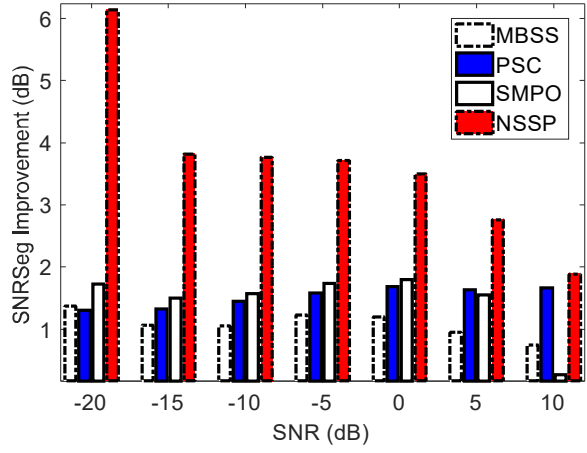

Figure 4: SNRSeg improvement for different methods in street noise.

Table 2: PESQ for different methods

| SNR(dB) | MBSS | PSC | SMPO | NSSP |
|---|---|---|---|---|
| -20 | 1.15 | 1.16 | 1.35 | 1.30 |
| -15 | 1.37 | 1.23 | 1.47 | 1.51 |
| -10 | 1.51 | 1.32 | 1.65 | 1.65 |
| -5 | 1.69 | 1.43 | 1.77 | 1.83 |
| 0 | 2.07 | 1.69 | 1.89 | 1.74 |
| 5 | 2.38 | 1.93 | 2.57 | 2.45 |
| 10 | 2.60 | 2.14 | 2.78 | 2.69 |

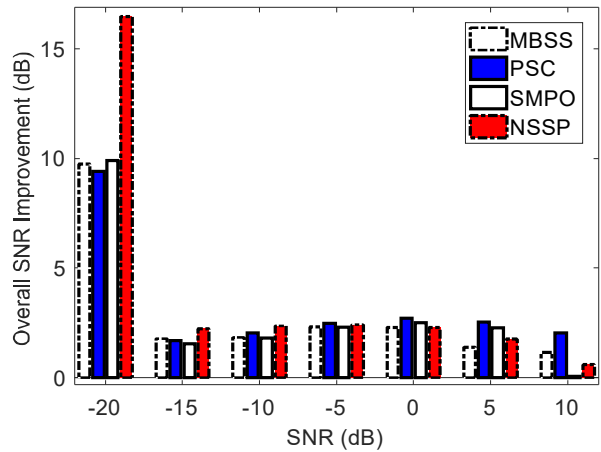

Figure 5: Overall SNR improvement for different methods in street noise.

PESQ values for different methods for all the SNR levels for street noise-corrupted speech are shown in Table 2. For higher SNR as 10 dB, we see that all the methods provide better PESQ. But with the decrement of SNR, PESQ values for all the cases start to fall. The proposed method provides very competitive PESQ values for all SNR levels in comparison to SMPO but performs better than other two competing methods. As PESQ value indicates the perceptual quality of the enhanced speech, this table proves that the proposed method provides better enhanced speech in street noise corrupted speech at high as well as low SNRs than MBSS and PSC.

*3.4.2. Results for speech signals with multi-talker babble boise*

SNRSeg improvement, overall SNR improvement and PESQ scores for speech signals corrupted with babble noise for MBSS, PSC, SMPO and NSSP are shown in Fig. 6, 7 and 8.

In Fig. 6, the performance of the NSSP is compared with performances of other methods at different levels of SNR in terms of SNRSeg improvement. From this figure, we see that the SNRSeg improvement in dB increases as SNR decreases for NSPP for all SNR levels. This is not true for MBSS, PSC and SMPO. Below SNR level of −10 dB, most of these three methods start to loose overall SNR improvement. At a low SNR of −20 dB, NSSP yields the highest SNRSeg improvement of more than 6 dB. Such larger values of SNRSeg improvement at a low level of SNR attest the capability of NSSP in producing enhanced speech with better quality for speech corrupted by babble noise- severely.

The overall SNR improvements of MBSS, PSC, SMPO and NSSP are shown in Fig. 7, where it is seen that NSSP provides an improvement of almost 18 dB at SNR level of −20 dB, which is significantly better than other methods. This trend continues upto 0 dB. After that NSSP provides competitive value in comparison to PSC and SMPO.

PESQ values for different methods are shown in Fig. 8 for noisy speech in babble noise. We see from this figure that although NSSP provides competitive PESQ scores in comparison to other methods for SNR levels of −20 to −10 dB, it provides higher PESQ scores for all other SNR levels.

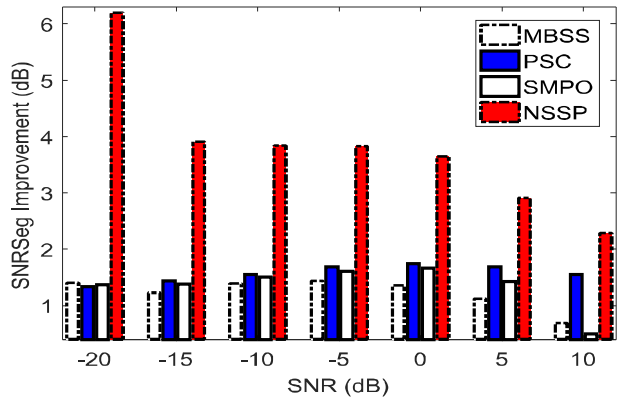

Figure 6: SNRSeg improvement for different methods in babble noise.

*3.5. Subjective Evaluation*

To evaluate the performance of the proposed method and other competing methods subjectively, we use two commonly used tools. The first one is the plot of the spectrograms of the outputs of all the methods and compare their performances in terms of preservation of harmonics and capability to remove noise.

The spectrograms of the clean speech, the noisy speech, and the enhanced speech signals obtained by using the proposed method and all other methods are presented in Fig. 9 for street noise corrupted speech at an SNR of 10 dB. It is obvious from the spectrograms that the proposed method preserves the harmonics significantly better than all the other competing methods. The noise is also reduced at every time point for the proposed method which attest our claim of better performance in terms of higher SNRSeg improvement, higher overall SNR improvement and higher PESQ values in objective evaluation. Another collection of spectrograms for the proposed method with other methods for speech signals corrupted with babble noise is shown in Fig. 10. This figure also attests that our proposed method has better performance in terms of harmonics' preservation and noise removal in presence of street noise.

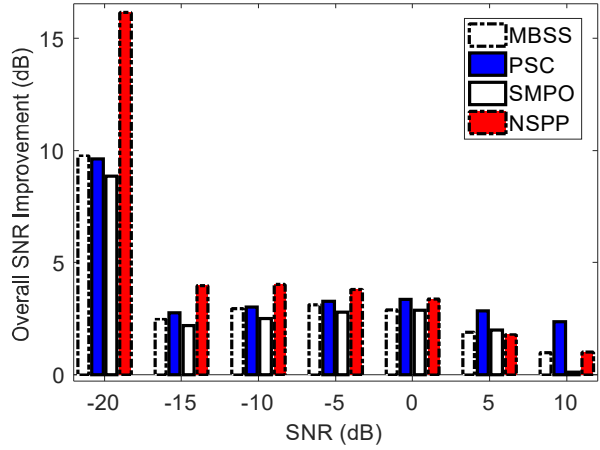

Figure 7: Overall SNR improvement for different methods in babble noise.

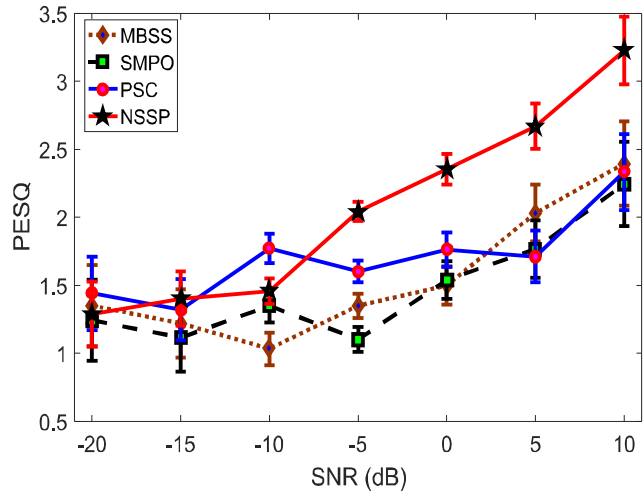

Figure 8: PESQ for different methods in babble noise.

The second tool we used for subjective evaluation of the proposed method and the competing methods is the formal listening tests. We add street and babble noises to all the thirty speech sentences of NOIZEUS database at −20 to 10 dB SNR levels and process them with all the competing methods. We allow ten listeners to listen to these enhanced speeches from these methods and evaluate them subjectively. Following [13] and [25], We use SIG, BAK and OVL scales on a range of 1 to 5. The detail of these scales and procedure of this listening test is discussed in [13]. More details on this testing methodology of listening test can be obtained from [26].

We show the mean scores of SIG, BAK, and OVRL scales for all the methods for speech signals corrupted with 10 dB street noise in Tables 3, 4, and 5 and for speech signals corrupted with 10 dB babble noise in Tables 6, 7, and 8. The higher values for the proposed method in comparison to other methods in these tables clearly attest that the proposed method is better than the competing methods in terms of lower signal distortion (higher SIG scores), efficient noise removal (higher BAK scores) and overall sound quality (higher OVL scores) for all SNR levels.

Table 3: Mean scores of SIG scale for different methods in presence of street noise at 10 dB

| Listener | MBSS | PSC | SMPO | NSSP |
|---|---|---|---|---|
| 1 | 4.2 | 3.5 | 4.0 | 4.1 |
| 2 | 3.8 | 3.4 | 3.8 | 3.9 |
| 3 | 4.0 | 3.4 | 4.1 | 4.3 |
| 4 | 4.1 | 3.9 | 4.2 | 4.6 |
| 5 | 3.2 | 3.3 | 3.9 | 4.5 |
| 6 | 3.4 | 3.2 | 4.6 | 3.4 |
| 7 | 3.5 | 3.4 | 3.8 | 4.3 |
| 8 | 3.6 | 3.2 | 4.1 | 4.3 |
| 9 | 3.4 | 3.2 | 4.5 | 3.4 |
| 10 | 3.7 | 3.9 | 4.8 | 4.5 |

Table 4: Mean scores of BAK scale for different methods in presence of street noise at 10 dB

| Listener | MBSS | PSC | SMPO | NSSP |
|---|---|---|---|---|
| 1 | 4.2 | 4.1 | 4.5 | 5.0 |
| 2 | 4.4 | 4.2 | 4.9 | 4.8 |
| 3 | 4.1 | 4.3 | 4.4 | 4.5 |
| 4 | 4.2 | 4.5 | 4.7 | 4.6 |
| 5 | 4.2 | 4.4 | 4.8 | 4.7 |
| 6 | 4.4 | 3.7 | 4.6 | 4.5 |
| 7 | 3.2 | 3.5 | 3.9 | 4.4 |
| 8 | 4.4 | 4.2 | 4.6 | 4.5 |
| 9 | 3.9 | 3.8 | 3.8 | 4.6 |
| 10 | 4.4 | 4.1 | 4.5 | 4.6 |

Table 5: Mean scores of OVL scale for different methods in presence of street noise at 10 dB

| Listener | MBSS | PSC | SMPO | NSSP |
|---|---|---|---|---|
| 1 | 4.2 | 2.9 | 4.0 | 4.3 |
| 2 | 3.8 | 3.8 | 3.9 | 3.9 |
| 3 | 4.6 | 3.5 | 4.0 | 4.4 |
| 4 | 4.4 | 3.5 | 4.2 | 4.3 |
| 5 | 3.5 | 3.4 | 3.8 | 4.4 |
| 6 | 4.1 | 3.3 | 3.6 | 4.6 |
| 7 | 3.2 | 3.2 | 3.8 | 4.7 |
| 8 | 4.5 | 3.8 | 3.7 | 4.5 |
| 9 | 4.4 | 3.9 | 3.9 | 4.4 |
| 10 | 4.2 | 3.4 | 3.9 | 4.8 |

Table 6: Mean scores of SIG scale for different methods in presence of babble noise at 10 dB

| Listener | MBSS | PSC | SMPO | NSSP |
|---|---|---|---|---|
| 1 | 4.0 | 3.6 | 4.0 | 4.3 |
| 2 | 3.9 | 3.3 | 3.9 | 3.9 |
| 3 | 4.0 | 3.9 | 4.0 | 4.4 |
| 4 | 4.2 | 3.4 | 4.2 | 4.7 |
| 5 | 3.8 | 3.2 | 3.8 | 4.4 |
| 6 | 3.6 | 2.9 | 3.6 | 3.9 |
| 7 | 3.8 | 3.8 | 3.8 | 4.4 |
| 8 | 3.6 | 3.4 | 3.6 | 4.3 |
| 9 | 3.9 | 3.5 | 3.9 | 3.9 |
| 10 | 3.8 | 3.7 | 3.8 | 3.9 |

Table 7: Mean scores of BAK scale for different methods in presence of babble noise at 10 dB

| Listener | MBSS | PSC | SMPO | NSSP |
|---|---|---|---|---|
| 1 | 4.5 | 4.0 | 4.5 | 4.6 |
| 2 | 4.9 | 4.3 | 4.9 | 4.4 |
| 3 | 4.4 | 4.2 | 4.4 | 4.7 |
| 4 | 4.7 | 4.4 | 4.7 | 4.8 |
| 5 | 4.8 | 4.2 | 4.8 | 4.8 |
| 6 | 4.6 | 3.9 | 4.6 | 4.7 |
| 7 | 3.9 | 3.8 | 3.9 | 4.6 |
| 8 | 4.6 | 4.4 | 4.6 | 4.5 |
| 9 | 3.9 | 3.5 | 3.9 | 4.5 |
| 10 | 4.8 | 4.7 | 4.8 | 4.6 |

Table 8: Mean scores of OVL scale for different methods in presence of babble noise at 10 dB

| Listener | MBSS | PSC | SMPO | NSSP |
|---|---|---|---|---|
| 1 | 4.1 | 2.9 | 4.0 | 4.4 |
| 2 | 3.9 | 3.4 | 3.8 | 3.8 |
| 3 | 4.5 | 3.4 | 4.1 | 4.3 |
| 4 | 4.1 | 3.3 | 4.2 | 4.3 |
| 5 | 3.8 | 3.2 | 3.9 | 4.5 |
| 6 | 4.4 | 3.7 | 4.6 | 4.6 |
| 7 | 3.9 | 3.4 | 3.8 | 4.5 |
| 8 | 4.0 | 3.6 | 4.1 | 4.2 |
| 9 | 4.4 | 3.1 | 4.5 | 4.7 |
| 10 | 4.5 | 3.1 | 4.8 | 4.9 |

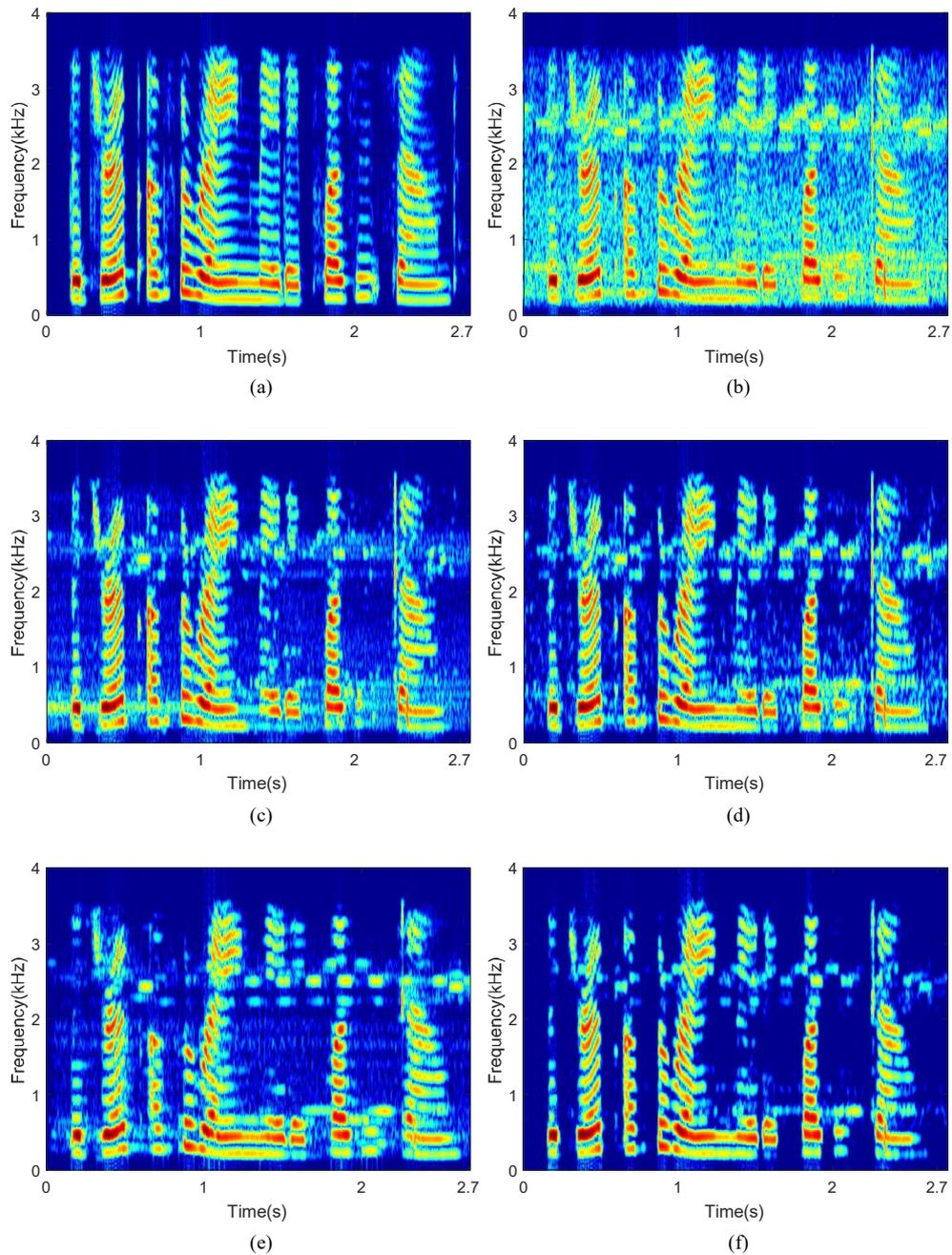

Figure 9: Spectrograms of (a) clean signal (b) noisy signal with 10 dB street noise; spectrograms of enhanced speech from (c) MBSS (d) PSC (e) SMPO (f) NSSP.

## 4. Conclusions

An improved speech enhancement method based on magnitude and phase compensation is presented in this paper for enhancement of noisy speech in adverse environment. Spectral subtraction is used in the first step for magnitude compensation depending on a new non-stationary noise estimation. In the second step, an SNR-dependent phase spectrum compensation is used to compensate the phase. For noisy speeches with medium to low levels of SNR, simulation results show that the proposed method yields consistently better results in the sense of higher segmental SNR improvement, overall SNR improvement, and output PESQ than those of the existing speech enhancement methods.

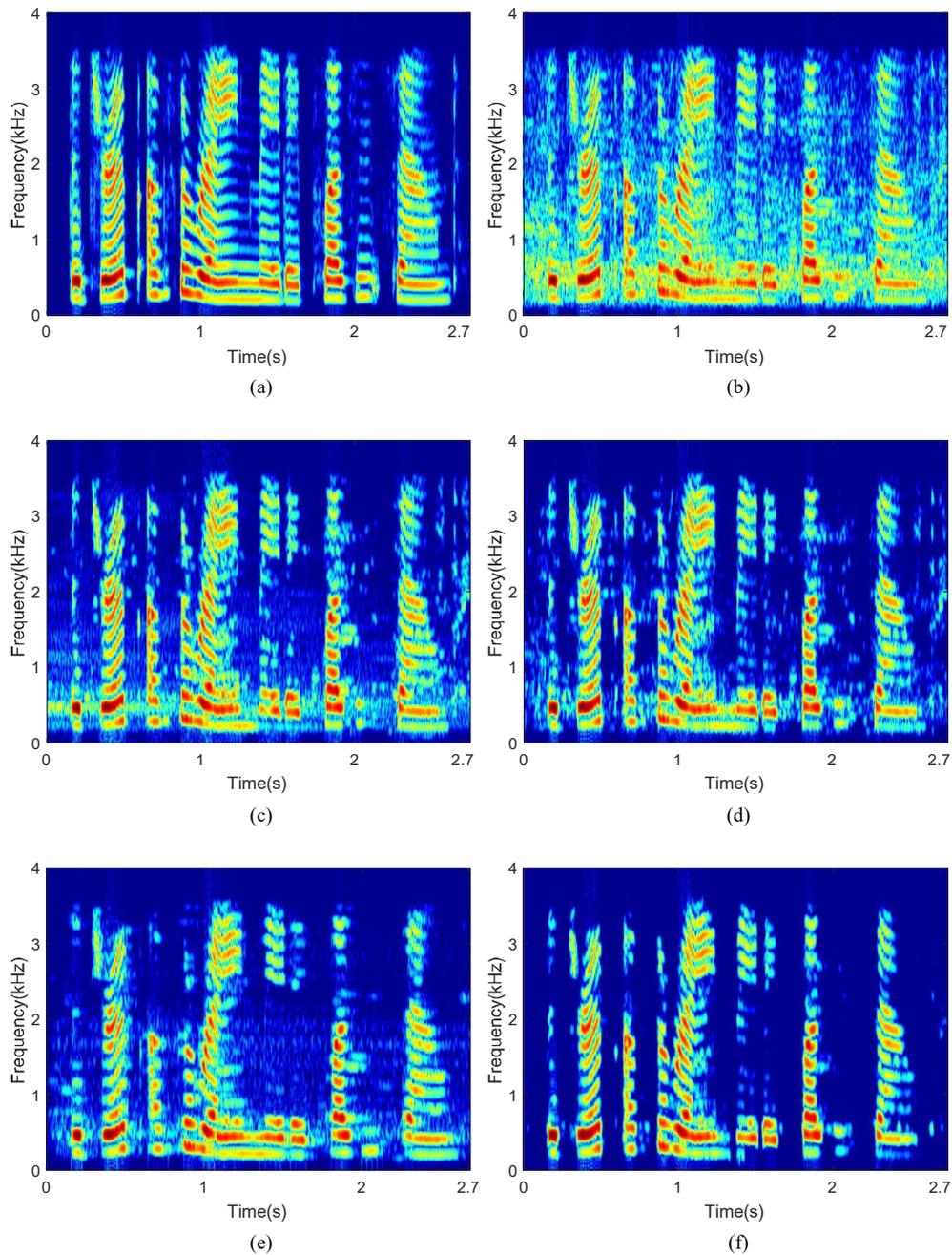

Figure 10: Spectrograms of (a) clean signal (b) noisy signal with 10 dB babble noise; spectrograms of enhanced speech from (c) MBSS (d) PSC (e) SMPO (f) NSSP.

[1] S. Boll, Suppression of acoustic noise in speech using spectral subtraction, IEEE Transactions on acoustics, speech, and signal processing 27 (2) (1979) 113–120.

[2] K. Yamashita, T. Shimamura, Nonstationary noise estimation using low-frequency regions for spectral subtraction, IEEE Signal processing letters 12 (6) (2005) 465–468.
[3] Y. Lu, P. C. Loizou, A geometric approach to spectral subtraction, Speech communication 50 (6) (2008) 453–466.
[4] M. T. Islam, C. Shahnaz, S. Fattah, Speech enhancement based on a modified spectral subtraction method, in: 2014 IEEE 57th International Midwest Symposium on Circuits and Systems (MWSCAS), IEEE, 2014, pp. 1085–1088.
[5] M. T. Islam, A. B. Hussain, K. T. Shahid, U. Saha, C. Shahnaz, Speech enhancement based on noise compensated magnitude spectrum, in: Informatics, Electronics Vision (ICIEV), 2014 International Conference on, IEEE, 2014, pp. 1–5.
[6] Y. Ephraim, H. L. Van Trees, A signal subspace approach for speech enhancement, IEEE Transactions on speech and audio processing 3 (4) (1995) 251–266.
[7] Y. Hu, P. C. Loizou, A generalized subspace approach for enhancing speech corrupted by colored noise, IEEE Transactions on Speech and Audio Processing 11 (4) (2003) 334–341.
[8] Y. Ephraim, D. Malah, Speech enhancement using a minimum mean-square error log-spectral amplitude estimator, IEEE Transactions on Acoustics, Speech, and Signal Processing 33 (2) (1985) 443–445.
[9] Y. Hu, P. C. Loizou, Subjective comparison and evaluation of speech enhancement algorithms, Speech communication 49 (7) (2007) 588–601.
[10] D. L. Donoho, De-noising by soft-thresholding, IEEE transactions on information theory 41 (3) (1995) 613–627.
[11] M. Bahoura, J. Rouat, Wavelet speech enhancement based on the teager energy operator, IEEE Signal Process. Lett. 8 (2001) 10–12.
[12] Y. Ghanbari, M. Mollaei, A new approach for speech enhancement based on the adaptive thresholding of the wavelet packets, Speech Commun. 48 (2006) 927–940.
[13] M. T. Islam, C. Shahnaz, W.-P. Zhu, M. O. Ahmad, Speech enhancement based on student modeling of teager energy operated perceptual wavelet packet coefficients and a custom thresholding function, IEEE/ACM Transactions on Audio, Speech, and Language Processing 23 (11) (2015) 1800–1811.
[14] N. Ma, M. Bouchard, R. A. Goubran, Speech enhancement using a masking threshold constrained kalman filter and its heuristic implementations, IEEE Transactions on Audio, Speech, and Language Processing 14 (1) (2006) 19–32.
[15] K. Paliwal, Estimation of noise variance from the noisy ar signal and its application in speech enhancement, IEEE Transactions on Acoustics, Speech, and Signal Processing 36 (2) (1988) 292–294.
[16] R. Martin, Noise power spectral density estimation based on optimal smoothing and minimum statistics, IEEE Transactions on speech and audio processing 9 (5) (2001) 504–512.
[17] J. Yamauchi, T. Shimamura, Noise estimation using high frequency regions for spectral subtraction, IEICE TRANSACTIONS on Fundamentals of Electronics, Communications and Computer Sciences 85 (3) (2002) 723–727.
[18] K. Wójcicki, M. Milacic, A. Stark, J. Lyons, K. Paliwal, Exploiting conjugate symmetry of the short-time fourier spectrum for speech enhancement, IEEE Signal processing letters 15 (2008) 461–464.
[19] A. P. Stark, K. K. Wójcicki, J. G. Lyons, K. K. Paliwal, K. K. Paliwal, Noise driven short-time phase spectrum compensation procedure for speech enhancement., in: INTERSPEECH, 2008, pp. 549–552.
[20] M. T. Islam, C. Shahnaz, Speech enhancement based on noise-compensated phase spectrum, in: Electrical Engineering and Information & Communication Technology (ICEEICT), 2014 International Conference on, IEEE, 2014, pp. 1–5.
[21] D. O'shaughnessy, Speech communication: human and machine, Universities press, 1987.
[22] S. Kamath, P. Loizou, A multi-band spectral subtraction method for enhancing speech corrupted by colored noise, in: IEEE international conference on acoustics speech and signal processing, Vol. 4, Citeseer, 2002, pp. 4164–4164.
[23] Y. Lu, P. C. Loizou, Estimators of the magnitude-squared spectrum and methods for incorporating snr uncertainty, IEEE transactions on audio, speech, and language processing 19 (5) (2011) 1123–1137.
[24] Y. Hu, P. C. Loizou, Evaluation of objective quality measures for speech enhancement, IEEE Transactions on audio, speech, and language processing 16 (1) (2008) 229–238.
[25] Y. Hu, P. Loizou, Subjective comparison and evaluation of speech enhancement algorithms, Speech Commun. 49 (2007) 588–601.
[26] ITU, P835 IT: subjective test methodology for evaluating speech communication systems that include noise suppression algorithms., ITU-T Recommendation (ITU, Geneva) (2003) 835.